\documentclass[twocolumn]{aastex631}

\received{2023 March 15}
\revised{2023 May 5}
\accepted{2023 May 23}

\begin{document}
\title{Modeling the chemical enrichment history of the Bulge Fossil Fragment Terzan~5}

\correspondingauthor{Donatella Romano}
\email{donatella.romano@inaf.it}

\author[0000-0002-0845-6171]{Donatella Romano}
\affiliation{INAF, Osservatorio di Astrofisica e Scienza dello Spazio, Via Gobetti 93/3, 40129 Bologna, Italy}

\author[0000-0002-2165-8528]{Francesco R.~Ferraro}
\affiliation{Dipartimento di Fisica e Astronomia, Universit\`a di Bologna, Via Gobetti 93/2, 40129 Bologna, Italy}
\affiliation{INAF, Osservatorio di Astrofisica e Scienza dello Spazio, Via Gobetti 93/3, 40129 Bologna, Italy}

\author[0000-0002-6040-5849]{Livia Origlia}
\affiliation{INAF, Osservatorio di Astrofisica e Scienza dello Spazio, Via Gobetti 93/3, 40129 Bologna, Italy}

\author[0000-0001-5839-0302]{Simon Portegies Zwart}
\affiliation{Leiden Observatory, Leiden University, PO Box 9513, 2300 RA Leiden, The Netherlands}

\author[0000-0001-5613-4938]{Barbara Lanzoni}
\affiliation{Dipartimento di Fisica e Astronomia, Universit\`a di Bologna, Via Gobetti 93/2, 40129 Bologna, Italy}
\affiliation{INAF, Osservatorio di Astrofisica e Scienza dello Spazio, Via Gobetti 93/3, 40129 Bologna, Italy}

\author{Chiara Crociati}
\affiliation{Dipartimento di Fisica e Astronomia, Universit\`a di Bologna, Via Gobetti 93/2, 40129 Bologna, Italy}
\affiliation{INAF, Osservatorio di Astrofisica e Scienza dello Spazio, Via Gobetti 93/3, 40129 Bologna, Italy}

\author[0000-0001-8892-4301]{Davide Massari}
\affiliation{INAF, Osservatorio di Astrofisica e Scienza dello Spazio, Via Gobetti 93/3, 40129 Bologna, Italy}

\author[0000-0003-4237-4601]{Emanuele Dalessandro}
\affiliation{INAF, Osservatorio di Astrofisica e Scienza dello Spazio, Via Gobetti 93/3, 40129 Bologna, Italy}

\author[0000-0001-9158-8580]{Alessio Mucciarelli}
\affiliation{Dipartimento di Fisica e Astronomia, Universit\`a di Bologna, Via Gobetti 93/2, 40129 Bologna, Italy}
\affiliation{INAF, Osservatorio di Astrofisica e Scienza dello Spazio, Via Gobetti 93/3, 40129 Bologna, Italy}

\author[0000-0003-0427-8387]{R.~Michael Rich}
\affiliation{Department of Physics and Astronomy, University of California, 90095 Los Angeles, CA, USA}

\author[0000-0002-6175-0871]{Francesco Calura}
\affiliation{INAF, Osservatorio di Astrofisica e Scienza dello Spazio, Via Gobetti 93/3, 40129 Bologna, Italy}

\author[0000-0001-7067-2302]{Francesca Matteucci}
\affiliation{Sezione di Astronomia, Dipartimento di Fisica, Universit\`a di Trieste, Via Tiepolo 11, 34143 Trieste, Italy}
\affiliation{INAF, Osservatorio Astronomico di Trieste, Via Tiepolo 11, 34143 Trieste, Italy}
\affiliation{INFN, Sezione di Trieste, Via Valerio 2, 34127 Trieste, Italy}




\begin{abstract}
Terzan~5 is a heavily obscured stellar system located in the inner Galaxy. It has been postulated to be a stellar relic, a Bulge Fossil Fragment witnessing the complex history of the assembly of the Milky Way bulge. In this paper, we follow the chemical enrichment of a set of putative progenitors of Terzan~5 to assess whether the chemical properties of this cluster fit within a formation scenario in which it is the remnant of a primordial building block of the bulge. We can explain the metallicity distribution function and the runs of different element-to-iron abundance ratios as functions of [Fe/H] derived from optical-infrared spectroscopy of giant stars in Terzan~5, by assuming that the cluster experienced two major star formation bursts separated by a long quiescent phase. We further predict that the most metal-rich stars in Terzan~5 are moderately He-enhanced and a large spread of He abundances in the cluster, $Y \simeq 0.26$--0.335. We conclude that current observations fit within a formation scenario in which Terzan~5 originated from a pristine, or slightly metal-enriched, gas clump about one order of magnitude more massive than its present-day mass. Losses of gas and stars played a major role in shaping Terzan~5 the way we see it now. The iron content of the youngest stellar population is better explained if the white dwarfs that give rise to type Ia supernovae (the main Fe factories) sink towards the cluster center, rather than being stripped by the strong tidal forces exerted by the Milky Way in the outer regions.
\end{abstract}




\section{Introduction}
\label{sec:intro}

The formation and evolution of bulges \citep[encompassing \emph{classical} and \emph{pseudobulges},][]{korm04} in massive spiral galaxies can be driven by several mechanisms, including violent, early dissipative collapse of gas \citep{egge62}, mergers \citep{spri05}, secular evolution of dynamically unstable discs \citep{comb90}, coalescence of giant star-forming clumps \citep{imme04,elme08}, or a combination thereof \citep[see][for a review]{atha05}.

In our Galaxy the bulge formation process is likely a motley growth, leading to a composite system. Although the central regions of the Milky Way (MW) are notoriously challenging to observe due to crowding and variable extinction on small spatial scales that severely hamper an unbiased view \citep{baad46,gonz12,nogu21}, it can be plainly argued that the Galactic bulge is a complex environment, where different stellar populations coexist \citep[][see also \citealt{atha17}]{bens13,roja14,barb18,hort21,quei21,nieu23}.

In the past decades, deep photometry and high-resolution spectroscopy of large samples of individual stars have provided key information on the ages, kinematics and chemical composition of the stellar populations inhabiting the MW bulge. The old ages inferred from colour-magnitude diagrams (CMDs) of different globular clusters (GCs) and fields in the bulge  \citep{orto95,zocc03,clar08,vale13,bica16,sara19,suro19} rule out an extended period of star formation. Yet, some stars with ages $\le 5$~Gyr are present \citep[e.g.,][]{bens13,bens17,catc16,schu17}. The significance of this young population is the subject of ongoing debate \citep{hayw16,renz18,saha19,rich20}. \citet{hass20} note that the likelihood of ending up with some relatively young stars in a bulge sample depends on the metallicity and height above the plane probed by the observations. Indeed, younger stars are found among the metal-richest ones and closer to the Galactic plane.

Field stars in the bulge cover a broad metallicity range, $-1.9 \le$ [Fe/H] $\le +0.6$ \citep{bens13}, with a few extremely metal-poor stars at [Fe/H]~$< -3$ \citep{howe15}. Metal-poor and metal-rich stars present distinct kinematic properties, overall consistent with belonging, respectively, to an old spheroid (or thick disc) and a buckled bar \citep{hill11,babu16}. In a [$\alpha$/Fe]--[Fe/H] diagram, members of the old, metal-poor, pressure-supported spheroidal component dominate a well-defined upper sequence, while objects inhabiting the metal-rich, boxy/peanut X-shaped bar populate the lower sequence, which sets important constraints on the formation timescale of each component \citep{roja19}. The complexity of the stellar populations further shows up in the metallicity distribution function (MDF), which unveils a clear multimodality\footnote{Other works \citep{bens11,hill11,roja14,gonz15,zocc17,schu17} reveal a bimodality in the MDF, possibly due to smaller sample sizes and/or larger individual measurement errors.} with varying proportions of its main components in different fields of view \citep{bens13,roja20}.

In terms of chemical properties, GCs in the bulge behave as their halo and thick-disc counterparts. Their members show the characteristic anticorrelations among the abundances of light elements, while the abundances of heavy elements do not present internal variations and follow the average abundance patterns traced by field stars \citep{grat19}.

Clusters found in the inner Galaxy possibly had larger masses in the past. Likely, they have lost preferentially low-mass stars, owing to mass segregation and strong tidal forces \citep{vesp97,baum03}. These processes may eventually lead to the complete dissolution of the clusters \citep{port02}. Recently, \citet{minn18} have reported the first clear observational signature of bulge-crossing shocks for M\,62, one of the most massive MW GCs located in the bulge. Another important point is that in $N$-body calculations the stellar remnants strongly concentrate towards the cluster core \citep[e.g.,][]{baum03}.

A few stellar systems in the MW bulge deserve special attention. In fact, up to now, two stellar systems in the MW bulge with the appearance of massive GCs, namely, Terzan~5 and Liller~1, have been found to harbor multiple sub-populations with large differences in age and in iron content \citep{ferr09,ferr16,ferr21}. The modeling of the star formation history of Liller~1 suggests that this stellar system has been forming stars over its entire lifetime, with three main star formation episodes -- the oldest occurring some 12--13~Gyr ago and the most recent one occurring only 1--3 Gyr ago -- and some low-level activity in between \citep{dale22}. Indeed, the first spectroscopic screening of Liller~1 confirms the presence of multi-iron sub-populations \citep{croc23}.

The case of Terzan~5, which is the target of this study, is better observationally constrained. Terzan~5 is a dense conglomeration of stars formerly classified as a GC, which is now known to harbor two major populations of stars, one old (age~= $12 \pm 1$~Gyr) and relatively metal-poor, with [Fe/H]$_{\mathrm{peak}} = -0.25 \pm 0.07$ and [$\alpha$/Fe]~$= +0.34 \pm 0.06$, the other younger (age~= $4.5 \pm 0.5$~Gyr) and metal-rich, with [Fe/H]$_{\mathrm{peak}} = +0.27 \pm 0.04$ and [$\alpha$/Fe]~$= +0.03 \pm 0.04$ \citep{ferr09,ferr16,orig11,orig13,orig19}. A third, minor component is also detected, peaking at [Fe/H]$_{\mathrm{peak}} =-0.79 \pm 0.04$ and with [$\alpha$/Fe]~$= +0.36 \pm 0.04$ \citep{orig13,mass14}. Although an additional ultra metal-rich component at [Fe/H]~$= +0.5$ could be possibly present, the statistics is still too low to firmly confirm its existence. The CMD analysis \citep{ferr09,ferr16} and the reconstructed star formation history (Crociati et al., in preparation) clearly point to two major star formation events, separated by a long quiescent -- or low-level activity -- period. These characteristics definitely peg Terzan~5 to a more massive progenitor that was able to retain its supernova (SN) ejecta and self-enrich. Such an object would be seen as a giant star-forming clump at high redshift \citep{bour08,elme08}. Interestingly, the peaks in the MDF of Terzan~5 track closely those in the MDF of the general bulge population, [Fe/H]~$= +0.32$, $-0.17$, and $-0.66$ \citep{roja20}. This would suggest a synchronous formation \citep{pfla09,mcke18}, rather than the accretion of a satellite previously evolved in isolation \citep[a scenario that, instead, explains rather well the dynamical and chemical properties of another anomalous GC-like system, $\omega$ Centauri,][]{bekk03,roma07}. Indeed, the orbit of Terzan~5 suggests an \emph{in-situ} origin \citep{mass15,baum19}. Is it hence possible that a non-negligible fraction of bulge stars formed within gaseous cocoons that self-enriched and then disrupted, leaving behind a few compact survivors? Could Terzan~5 and Liller~1 be the remnants of these building blocks? In order to answer these questions, here we explore possible evolutionary paths yielding to a stellar system with structure and chemical properties compatible with the present-day configuration of Terzan~5.

The layout of the paper is as follows. The chemical evolution model is described in Sect.~\ref{sec:model}. The results are compared to the available observations in Sect.~\ref{sec:res}. Predictions still awaiting the test of future observations are presented in Sect.~\ref{sec:disc} and discussed in light of existing scenarios for the origin of Terzan~5. Our conclusions are drawn in Sect.~\ref{sec:disc}.

\section{Chemical evolution model}
\label{sec:model}

We follow the evolution of the abundances of several chemical elements representative of different nucleosynthesis channels (light, $\alpha$, Fe-peak, and neutron-capture elements) in the interstellar medium (ISM) of putative precursors of Terzan~5. To this end, we use a single-zone numerical model \citep{roma13,roma15} that solves the classical set of equations of chemical evolution \citep{tins80,matt12,matt21}.

We do not implement any detailed treatment of the gas and star dynamics in our model. Nonetheless, we are able to discuss some dynamical effects, by putting our findings into the broader context provided by $N$-body simulations of star cluster evolution (see Sects.~\ref{sec:res} and \ref{sec:disc}).

\subsection{Basic assumptions}
\label{sec:basic}

Raw material for star formation is provided by gas cooling within a giant cloud \citep[for a thorough discussion of gas accretion modes in galaxies see][]{sanc14}. We consider either pristine gas with primordial chemical composition ($Z_\mathrm{in} = 0$) or pre-enriched gas with [Fe/H]~$= -1$ ($Z_\mathrm{in} \simeq 0.005$). The rate of gas infall is parameterised as
\begin{equation}
\frac{\mathrm{d}M_\mathrm{in}}{\mathrm{d}t} \propto \mathrm{e}^{-t/\tau},
\label{eq:in}
\end{equation}
where $M_\mathrm{in}$ is the initial mass of the cloud, and the e-folding time, $\tau$, is a free parameter of the model.

The cold gas is turned into stars following a phenomenological law \citep{schm59,kenn98},
\begin{equation}
\psi(t) = \nu M_\mathrm{gas}^k(t),
\label{eq:sfr}
\end{equation}
where the star formation efficiency, $\nu$, is a free parameter of the model (set constant in time) and $k = 1$.

The multi-peaked MDF and CMD morphology of Terzan~5 suggest that the star formation in the proto-cluster was not a smooth, continuous process, but rather proceeded through distinct bursts \citep{ferr16}, possibly interspersed with low-level activity, as is the case for Liller~1 \citep{dale22}. It is important to keep in mind that, while the multi-peaked MDF of Terzan~5 clearly indicates the existence of multi-iron components, it cannot be used to weight precisely each sub-population. More stringent constraints to the sizes of the two major populations (the metal-poor one peaking at [Fe/H]$_{\mathrm{peak}} \simeq -0.3$, generated about 12~Gyr ago, and the metal-rich one peaking at [Fe/H]$_{\mathrm{peak}} \simeq +0.3$, formed 4.5~Gyr ago) can be derived from photometric properties. In fact, the star counts in the two detected red clumps \citep[see][]{ferr09} led to the estimate that approximately 38\% of the current stellar population of Terzan~5 (namely, $7.5 \times 10^5$~M$_\odot$ of stars) originated from the youngest burst \citep{lanz10}. Regarding the relative ages of the two sub-populations, it is worth emphasizing that the analysis of the main-sequence turnoff region \citep{ferr16} has definitely removed any age-helium degeneracy \citep[see ][for a thorough discussion of this problem]{dant10,nata12}. We use the above observational inferences to constrain the star formation history of our models.

The initial masses of newborn stars are distributed according to a \citet{krou01} stellar initial mass function (IMF). In this work, the IMF is normalized to unity over the 0.1--100~M$_\odot$ mass range.

In some models we also take gas outflows and/or stellar stripping into account. Energetic feedback from SNe, ram pressure and tidal stripping, in fact, may get rid of a large fraction of the gas left over from the star formation process. Clusters born in the inner Galaxy also have their low-mass stars first pushed in the outskirts by mass segregation, then peeled off by strong tidal forces, with complete dissolution of the cluster being a possible outcome \citep[see][]{port02}.

\subsection{Nucleosynthesis prescriptions}
\label{sec:nuc}
The stellar nucleosynthesis prescriptions are at the core of any chemical evolution model. In this study we adopt stellar yields that have been successfully tested against large chemical abundance datasets for the MW in previous work \citep{roma10,roma17,roma19}.

More precisely, the stellar yields are taken from \citet{kara10} for low- and intermediate-mass stars, \citet{doh14a,doh14b} for super-AGB stars, \citet{nomo13} for massive stars, and \citet{iwam99} for type Ia SNe (SNeIa, thermonuclear explosions of white dwarfs in binary systems that completely disrupt the parent system). Stars more massive than 20~M$_\odot$ may explode as core-collapse SNe (CCSNe), releasing energies of the order of $10^{51}$ ergs, or as hypernovae (HNe), releasing energies an order of magnitude higher. The HN fraction may vary (for instance, as a function of the metallicity of the system). To encompass all possible cases, we run both models in which all massive stars explode as CCSNe and models in which all stars with initial masses above 20~M$_\odot$ explode as HNe. The yield tables are linearly interpolated as a function of the initial stellar mass and metallicity, and extrapolated to the 40--100~M$_\odot$ regime (not covered by full stellar nucleosynthesis calculations in \citealt{nomo13}) by keeping the yield constant.

While the number of CCSNe/HNe that explode in a system is plain set by the adopted star formation history and IMF, the number of expected SNIa explosions may depend on additional factors. For instance, orbital hardening and exchange interactions may lead to enhanced SNIa rates in star clusters relative to the field \citep{shar02,shar06}. In classic chemical evolution models it is customary to assume that a constant fraction of the stellar mass in the range 3--16~M$_\odot$ enters the formation of binary systems that give rise to SNIa explosions \citep[see, e.g.,][]{matt01}. However, this fraction might be time and/or environment dependent. We discuss this issue in Sect.~\ref{sec:res}.

\section{Results}
\label{sec:res}

\begin{table}
\setlength{\tabcolsep}{2.2pt}
\begin{center}
\caption{Model parameters and final stellar mass of the single-burst models\label{tab:mod-s}}
\begin{tabular}{@{}cccccccc@{}}
\hline
\hline
Model & $M_\mathrm{in}$ & $Z_\mathrm{in}$ & $\tau$ & $\Delta t_\mathrm{SF}$ & $\nu$            & stripped  & $M_\mathrm{stars}$ \\
      & (M$_\odot$)     &                 & (Gyr)  & (Gyr)                  & (Gyr$^{-1}$)     & stars     & (M$_\odot$)         \\
      &                 &                 &        &                        &                  & (\%)      &                     \\
(1)   & (2)             & (3)             & (4)    & (5)                    & (6)              & (7)       & (8)                 \\
\hline
{\footnotesize H01}   & $10^8$ & 0     & 0.5 & 1.0 & 0.27 & 0  & $10^7$            \\
{\footnotesize H02}   & $10^8$ & 0     & 1.0 & 1.0 & 0.31 & 0  & $10^7$            \\
{\footnotesize H03}   & $10^8$ & 0     & 2.0 & 1.0 & 0.44 & 0  & $10^7$            \\
{\footnotesize H04}   & $10^8$ & 0     & 3.0 & 1.0 & 0.59 & 0  & $10^7$            \\
{\footnotesize H05}   & $10^8$ & 0     & 4.0 & 1.0 & 0.74 & 0  & $10^7$            \\
{\footnotesize H06}   & $10^8$ & 0     & 5.0 & 1.0 & 0.89 & 0  & $10^7$            \\
{\footnotesize H07}   & $10^8$ & 0     & 6.0 & 1.0 & 1.03 & 0  & $10^7$            \\
{\footnotesize H08}   & $10^8$ & 0     & 7.0 & 1.0 & 1.16 & 0  & $10^7$            \\
{\footnotesize H08s}  & $10^8$ & 0     & 7.0 & 1.0 & 1.16 & 56 & $4.4$$ \times$$10^6$ \\
\tableline
{\footnotesize L01}   & $10^7$ & 0.005 & 0.5 & 1.0 & 0.27 & 0  & $10^6$            \\
{\footnotesize L02}   & $10^7$ & 0.005 & 1.0 & 1.0 & 0.31 & 0  & $10^6$            \\
{\footnotesize L03}   & $10^7$ & 0.005 & 2.0 & 1.0 & 0.44 & 0  & $10^6$            \\
{\footnotesize L04}   & $10^7$ & 0.005 & 3.0 & 1.0 & 0.59 & 0  & $10^6$            \\
{\footnotesize L05}   & $10^7$ & 0.005 & 4.0 & 1.0 & 0.74 & 0  & $10^6$            \\
{\footnotesize L06}   & $10^7$ & 0.005 & 5.0 & 1.0 & 0.89 & 0  & $10^6$            \\
{\footnotesize L06s}  & $10^7$ & 0.005 & 5.0 & 1.0 & 0.89 & 47 & $5.3$$\times$$10^5$ \\
{\footnotesize L07}   & $10^7$ & 0.005 & 6.0 & 1.0 & 1.03 & 0  & $10^6$            \\
{\footnotesize L08}   & $10^7$ & 0.005 & 7.0 & 1.0 & 1.16 & 0  & $10^6$            \\
\tableline
{\footnotesize T01}   & $10^7$ & 0.005 & 5.0 & 0.1 & 10.0 & 0  & $10^6$            \\
{\footnotesize T01s}  & $10^7$ & 0.005 & 5.0 & 0.1 & 10.0 & 67 & $3.3$$\times$$10^5$ \\
{\footnotesize T02}   & $10^7$ & 0.005 & 5.0 & 0.5 & 1.81 & 0  & $10^6$            \\
{\footnotesize T03}   & $10^7$ & 0.005 & 5.0 & 2.0 & 0.43 & 0  & $10^6$            \\
{\footnotesize T04}   & $10^7$ & 0.005 & 7.0 & 0.1 & 14.2 & 0  & $10^6$            \\
{\footnotesize T04s}  & $10^7$ & 0.005 & 7.0 & 0.1 & 14.2 & 52 & $4.8$$\times$$10^5$ \\
\hline
\end{tabular}
\end{center}
\tablecomments{This includes (1) model name, (2) total initial mass and (3) metallicity of the infalling gas, (4) infall time scale, (5) duration of the star formation, (6) star formation efficiency, (7) percentage of stripped stars (if any), and (8) present-day stellar mass of the system.}
\end{table}

In this section, we present the results of different sets of models, in which we vary in turn: the mass and initial chemical composition of the gaseous clump out of which Terzan~5 emerges, the infall timescale, the number and duration of the star formation episodes, and the star formation efficiency. Since there is little doubt that the mass of Terzan~5 was larger in the past than what is observed today, $M_\mathrm{stars} = (2 \pm 1) \times 10^6$~M$_\odot$ \citep{lanz10}, and motivated by dynamical arguments \citep{baum08}, we assume that the initial gas cloud mass is in the range $10^7$--$10^8$~M$_\odot$. We build models that produce a final stellar mass in the range $10^6$--$10^7$~M$_\odot$, and study the effects of gas and stellar stripping. If stellar stripping is ineffective, the lower limit to the stellar mass range set above is consistent with the lowest value permitted by the observations \citep{lanz10}, while the upper limit allows for a high stripping efficiency.

Firstly, we develop a series of exploratory models to analyse \emph{qualitatively} the outcome of an early star formation burst that builds up the majority of the stellar populations in Terzan~5 (Sects.~\ref{sec:high}--\ref{sec:delta}). Then, we add the fringe star formation episode that originates the metal-rich, younger stars, paying attention to reproduce \emph{quantitatively} the fractions of metal-poor and metal-rich stars inferred from the red clump analysis (Sect.~\ref{sec:comp}). The values of the input parameters are listed in Table~\ref{tab:mod-s} (Table~\ref{tab:mod-d}) for the single-burst (double-burst) models.

\subsection{A pristine, massive parent cloud}
\label{sec:high}

We start our analysis by considering a $10^8$~M$_\odot$ gas cloud with primordial chemical composition ($Z_\mathrm{in} = 0$). For the moment, we concentrate on the oldest stellar populations and forget about the young, metal-rich stars. We keep the duration of the ancient (age~$= 12$~Gyr) star formation episode that generates the old stars fixed to $\Delta t_\mathrm{SF} = 1$~Gyr and vary the e-folding time in Eq.~(\ref{eq:in}) and the star formation efficiency in Eq.~(\ref{eq:sfr}) so as to obtain the same present-day stellar mass ($M_\mathrm{stars} = 10^7$ M$_\odot$) for different choices of the ($\tau$, $\nu$) parameter couple. The adopted values of the model parameters are reported in Table~\ref{tab:mod-s} (models from {\footnotesize H01} to {\footnotesize H08}).

\begin{figure}
\centering
\includegraphics[width=21pc]{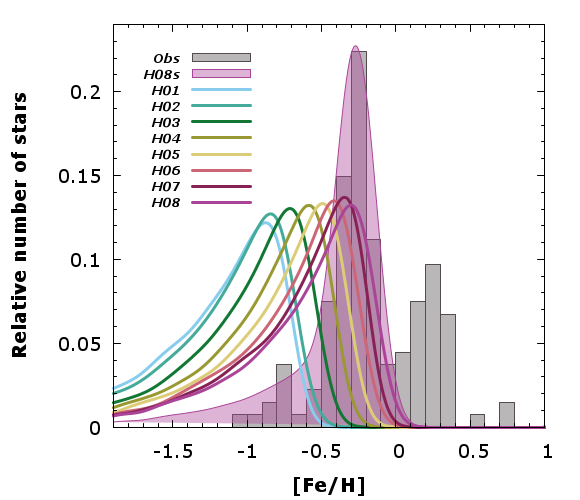}
\caption{Observed \citep[grey histogram,][]{mass14} and predicted (solid lines and shaded area) iron distributions of the stellar populations of Terzan~5. The theoretical distributions are convolved with a Gaussian of dispersion $\sigma = 0.1$ dex in order to take the observational errors into account. The shaded area refers to model~{\footnotesize H08s} that also takes into account stripping of low-metallicity stars (see text).}
\label{fig:mdf_high}
\end{figure}

\begin{figure}
\centering
\includegraphics[width=21pc]{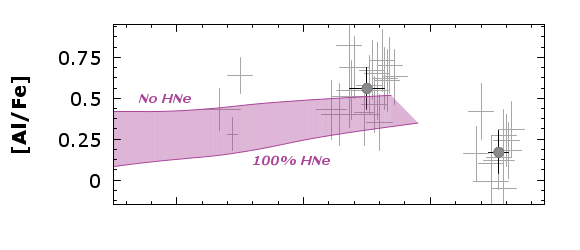}\vspace{-0.72cm}
\includegraphics[width=21pc]{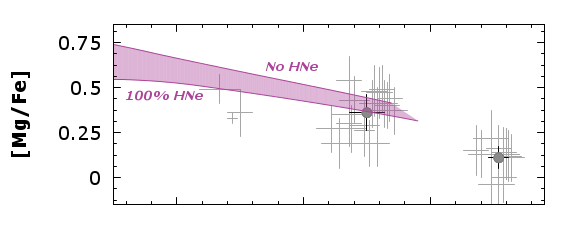}\vspace{-0.72cm}
\includegraphics[width=21pc]{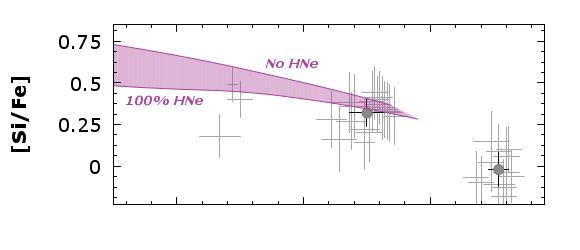}\vspace{-0.72cm}
\includegraphics[width=21pc]{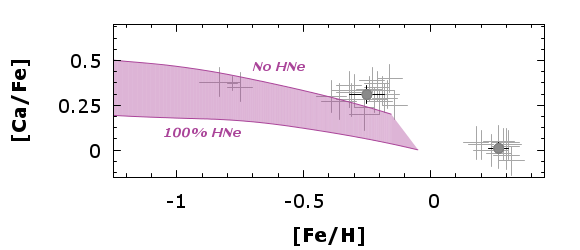}
\caption{\emph{From top to bottom:} runs of [Al/Fe], [Mg/Fe], [Si/Fe], and [Ca/Fe] as functions of [Fe/H] in Terzan~5 predicted by model~{\footnotesize H08s}, taking into account the uncertainties due to the fraction of stars that explode as HNe (shaded areas). Data for individual giants (crosses), along with average values for the different components (circles), with the exception of the metal-poor regime, where the sample amounts to only three objects, are from \citet{orig11,orig13}. Theoretical and observed abundance ratios are normalized to the solar values by \citet{magg22}.}
\label{fig:ele_high}
\end{figure}

\begin{figure}
\centering
\includegraphics[width=21pc]{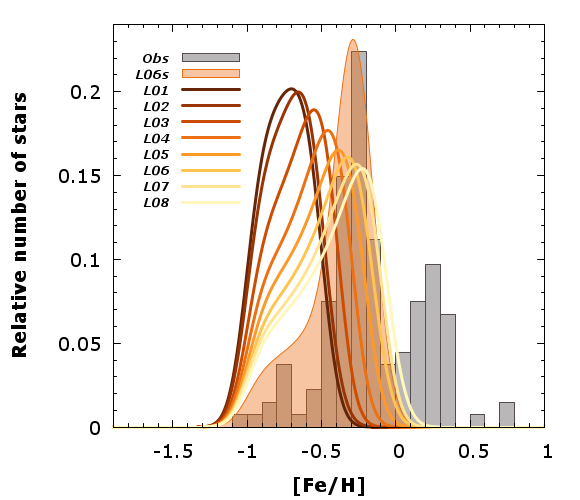}
\caption{Same as Fig.~\ref{fig:mdf_high}, but for models from {\footnotesize L01} through to {\footnotesize L08} (solid lines). The shaded area refers to model {\footnotesize L06s}, allowing for stellar stripping.}
\label{fig:mdf_low}
\end{figure}

\begin{figure}
\centering
\includegraphics[width=21pc]{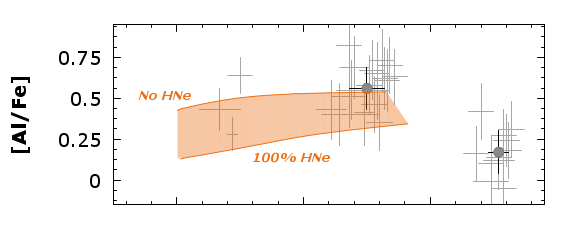}\vspace{-0.72cm}
\includegraphics[width=21pc]{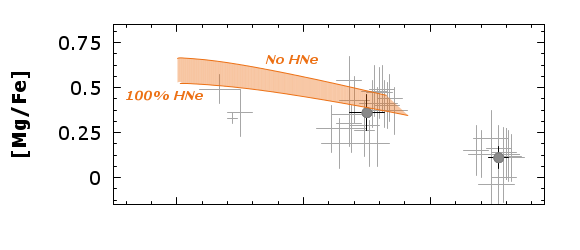}\vspace{-0.72cm}
\includegraphics[width=21pc]{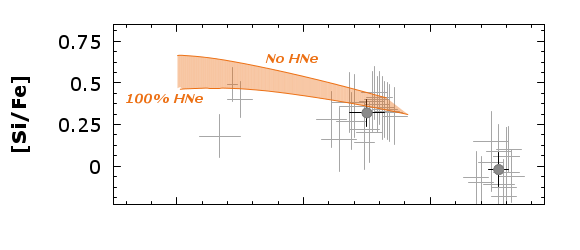}\vspace{-0.72cm}
\includegraphics[width=21pc]{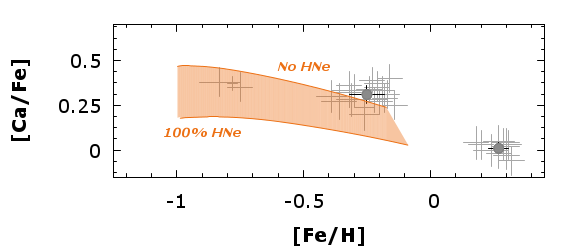}
\caption{Same as Fig.~\ref{fig:ele_high}, but for model {\footnotesize L06s}.}
\label{fig:ele_low}
\end{figure}

Clearly, the longer the infall time scale, the lower the cold gas mass available to form stars at early times and the higher the star formation efficiency required to end up with the desired total stellar mass. In Fig.~\ref{fig:mdf_high}, the predictions of models from {\footnotesize H01} through to {\footnotesize H08} (coloured lines) are compared to the observed iron distribution in Terzan~5 \citet[grey histogram,][]{mass14}. The longer the infall time scale and the higher the star formation efficiency, the more the peak of the theoretical MDF is skewed towards high metallicities, in better agreement with the observations.

We are facing, unsurprisingly, the well-known problem commonly referred to as the `G-dwarf problem', namely, a predicted excess of low-metallicity stars in systems that evolve as a closed box \citep{vdbe62,page75,mart00,gree21}. It is obvious that Terzan~5 does not evolve as an isolated system. Instead, its evolution is strongly influenced by the surroundings. In fact, a better fit to the observed MDF of Terzan~5 is obtained not only by assuming a somehow inefficient gas cooling ($\tau = 7$~Gyr for models~{\footnotesize H08} and {\footnotesize H08s} versus $\tau = 0.5$~Gyr for model~{\footnotesize H01}), possibly dictated by lively star formation activity in neighboring areas that keeps the gas warm, but also by accounting for the loss of a significant fraction of metal-poor stars due to mass segregation and tidal stripping\footnote{It is reasonable to expect that, as the star formation and the chemical enrichment proceed, cooling flows bring the enriched gas to the cluster center. Therefore, metal-rich stars form in the inner regions, while metal-poor stars are found mainly in the outer regions. For these reasons metal-poor, low-mass stars are more susceptible to stripping.} (model~{\footnotesize H08s}). The purple-shaded area in Fig.~\ref{fig:mdf_high} that best fits the data, in fact, refers to a situation in which 75\% of the stars with [Fe/H]~$< -0.4$ have been stripped, corresponding to a 56\% reduction of the total stellar mass. This leaves us with $M_\mathrm{stars} \simeq 4.4 \times 10^6$ M$_\odot$ at present, which is hardly compatible with the observed value for the old stellar component of Terzan~5, within the quoted uncertainty.

Figure~\ref{fig:ele_high} shows the behaviour of the abundance ratios of Al, Mg, Si, and Ca to Fe as a function of metallicity predicted by model~{\footnotesize H08s}. The lower (upper) envelope of the shaded area in each panel refers to the case in which HN nucleosynthesis is (is not) included, with all (no) stars above 20~M$_\odot$ exploding as HNe. The shaded area stands for intermediate figures. The agreement among model predictions and observations is satisfactory\footnote{We remind that the metal-rich sub-population is not reproduced by this model by construction.}, also taking into account the many uncertainties that still affect the stellar nucleosynthesis prescriptions \citep{roma10}.

\subsection{A pre-enriched, low-mass gas cocoon}
\label{sec:low}

\begin{figure}
\centering
\includegraphics[width=21pc]{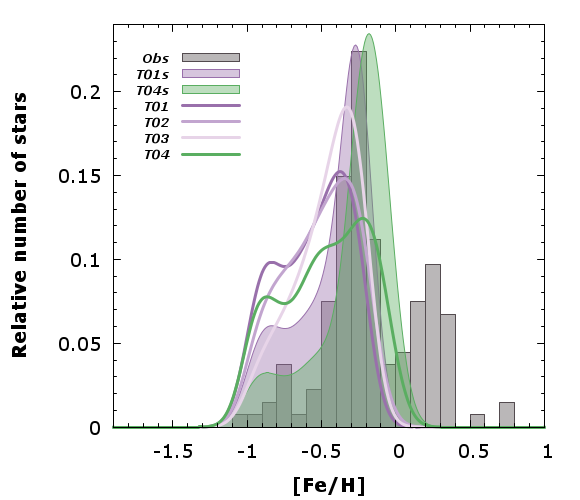}
\caption{Same as Fig.~\ref{fig:mdf_high}, but for models from {\footnotesize T01} through to {\footnotesize T04} (solid lines). The shaded areas refer to models including stellar stripping ({\footnotesize T01s}, {\footnotesize T04s}).}
\label{fig:mdf_delta}
\end{figure}

In this section, we discuss the results of a set of models (Table~\ref{tab:mod-s}, models from {\footnotesize L01} to {\footnotesize L08}) that share the same parameter values with the previous one, bar the total amount ($M_\mathrm{in} = 10^7$~M$_\odot$) and initial chemical composition ($Z_\mathrm{in} = 0.005$) of the gas. The final stellar mass is $M_\mathrm{stars} = 10^6$~M$_\odot$, with no synthetic stars created below [Fe/H]~$= -1$ by construction.

Similarly to what we have seen in the previous section, the theoretical MDF matches more closely the observed one if: (i) the gas is kept warm (due to strong stellar feedback in the surroundings?), so that the cold gas that fuels the star formation is accreted on long time scales and (ii) the system is deprived of a non-negligible fraction (here, 65\%) of its long-lived, low-mass stars with [Fe/H]~$\le -0.4$ dex (Fig.~\ref{fig:mdf_low}). After stripping, model~{\footnotesize L06s} ends with a total present-day stellar mass of $\sim 5.3 \times 10^5$~M$_\odot$, which is marginally compatible with the observational estimate (when also taking into account the increase in stellar mass that is expected as a consequence of the second, late star formation burst). The chemical properties of the sub-solar populations of Terzan~5 are well reproduced by model~{\small L06s} (Fig.~\ref{fig:ele_low}), which also matches well the position of the main peak of the MDF.

We note that starting from a slightly pre-enriched gas favours the appearance of a secondary, metal-poor MDF peak. This conclusion is totally unrelated to the assumed mass of the progenitor. What matters, in fact, are the relative proportions between the mass of the cold gas entering the star-forming regions (dictated by the choice of $\tau$) and the mass of the stars born in the system (dictated by the choice of $\nu$) at each time, and not their absolute values.

While our chemical evolution model can not put stringent constraints on the initial mass of the progenitor of Terzan~5 (though values as high as $M_\mathrm{in} = 10^8$~M$_\odot$ seem somehow disfavoured), it is pretty much consistent with the expectations from dynamical investigations. \citet{baum08} suggest that gas clouds with masses in excess of $10^7$~M$_\odot$ tend to retain their gas despite multiple SN events, which would explain the self-enrichment of the most massive clusters. However, it is worth emphasizing that such models are tailored to systems evolving in a much less dense environment than the one provided by the inner Galaxy in its infancy.

\subsection{Duration of the first star formation episode}
\label{sec:delta}

The analysis of the CMD of Terzan~5 allows an estimate of the ages of its stellar populations. \citet{ferr16} find that the sub-solar components are consistent with ages of $12 \pm 1$~Gyr. This means that the duration of the star formation episode originating the oldest stars in Terzan~5 is $\Delta t_\mathrm{SF} \le 2$~Gyr. In principle, it could be as short as few tens of Myr. The models discussed in the previous sections assume a fiducial value of $\Delta t_\mathrm{SF} = 1$~Gyr. In this section, we examine the effects of $\Delta t_\mathrm{SF}$ variations, from 100~Myr to 2~Gyr (see Table~\ref{tab:mod-s}, models~{\footnotesize T01}--{\footnotesize T04}).

Figure~\ref{fig:mdf_delta} illustrates the changes in the shape of the predicted MDF produced by variations of the duration of the main star formation episode compatible with the observed CMD (purple solid lines of progressively lighter shade for progressively longer star formation episodes). Since all the models have to produce the same final stellar mass, the shorter $\Delta t_\mathrm{SF}$, the higher the required star formation efficiency (Table~\ref{tab:mod-s}). A higher star formation rate at early times (model~{\footnotesize T01}, darkest purple line) results in a more pronounced secondary peak at [Fe/H]~$\simeq -0.8$, while such a feature completely disappears if the star formation lasts longer than 1~Gyr (model~{\footnotesize T03}, lightest purple line). On the other hand, the position of the main peak does not change much with changing $\Delta t_\mathrm{SF}$ -- it is rather dictated by the adopted value of $\tau$ (compare models~{\footnotesize T01} and {\footnotesize T04}, assuming $\tau = 5$ and 7~Gyr, respectively). Not differently from what previously found, a better agreement with the observed MDF of Terzan~5 is obtained by imposing that stellar stripping is removing the majority of the low-mass, metal-poor stars (80\% of all stars below [Fe/H]~$= -0.4$, models~{\footnotesize T01s} and {\footnotesize T04s}, shaded areas in Fig.~\ref{fig:mdf_delta}).

\begin{table*}
\setlength{\tabcolsep}{3.5pt}
\begin{center}
\caption{Model parameters and final stellar mass of the double-burst models\label{tab:mod-d}}
\begin{tabular}{@{}ccccccccccccc@{}}
\hline
\hline
Model & $M_\mathrm{in}$ & $Z_\mathrm{in}$ & $\tau$ & $\Delta t_\mathrm{SF_\mathrm{I}}$ & $\Delta t_\mathrm{SF_\mathrm{II}}$ & $\nu_\mathrm{I}$ & $\nu_\mathrm{II}$ & stripped stars & gas loss & SNIa precursors & $M_\mathrm{stars \, I}$ & $M_\mathrm{stars \, II}$ \\
      & (M$_\odot$)     &                 & (Gyr)  & (Gyr)                             & (Gyr)                              & (Gyr$^{-1}$)     & (Gyr$^{-1}$)                & (\%)           &          & retention     & (M$_\odot$)             & (M$_\odot$) \\
(1)   & (2)             & (3)             & (4)    & (5)                               & (6)                                & (7)              & (8)                               & (9)                       & (10)          & (11)     & (12)         & (13) \\
\hline
{\footnotesize S01s}  & $4$$\times$$10^7$ & 0.005 & 5.0 & 0.1 & 0.1 & 10.0 & 10.0 & 41 & no  & yes  & $1.3$$\times$$10^6$ & $3.0$$\times$$10^6$ \\
{\footnotesize S02s}  & $4$$\times$$10^7$ & 0.005 & 5.0 & 0.1 & 0.1 & 10.0 & 10.0 & 58 & yes & yes & $1.3$$\times$$10^6$ & $7.5$$\times$$10^5$ \\
{\footnotesize S03s}  & $4$$\times$$10^7$ & 0.005 & 5.0 & 0.1 & 0.1 & 10.0 & 10.0 & 58 & yes & no  & $1.3$$\times$$10^6$ & $7.5$$\times$$10^5$ \\
\hline
\end{tabular}
\end{center}
\tablecomments{This includes (1) model name, (2) total initial mass and (3) metallicity of the infalling gas, (4) infall time scale, (5) duration of the early and (6) late star formation episodes, (7) star formation efficiency of the first and (8) second star formation episode, (9) percentage of stripped stars (if any), (10) a flag that indicates if gas outflow/stripping is taken into account, (11) a flag that indicates whether SNIa progenitors are retained, and (12) the present-day stellar mass assembled in the first and (13) in the second starburst, respectively (their sum gives the current total mass of the system).}
\end{table*}

In principle, the duration of the star formation activity affects also the predicted trend of the elemental abundance ratios as a function of [Fe/H]. The longer $\Delta t_\mathrm{SF}$, in fact, the larger the amount of iron restored by SNeIa on long time scales that enters the formation of subsequent stellar generations, resulting in a prominent reduction of several element-to-iron abundance ratios at high metallicities (especially for those elements that are produced for the most part by massive stars). However, the request that the final stellar mass be the same for all models translates into the demand for greater star formation efficiencies when the star formation lasts less. This leads to a larger production of Fe from massive stars on short time scales in the models with quicker bursts, which accompanies the enhanced, fast making of all other massive star products. The net result is that different models reach similar final metallicities, while the element-to-iron ratios differences amount to 0.2~dex at most. The most effective ‘litmus test' of the duration of the star formation burst is, thus, the observed MDF of the system.

\subsection{The two-burst models}
\label{sec:comp}

The sub-solar metallicity components of Terzan~5 have chemical properties consistent with being born on a relatively short timescale ($\Delta t_\mathrm{SF_\mathrm{I}} = 100$~Myr) from a slightly metal-enriched gas cocoon significantly more massive (with a mass of a few $10^7$~M$_\odot$) than the present system. In order to reproduce the chemical properties of the youngest (age~$= 4.5$~Gyr) population of Terzan~5 that accounts for a non negligible fraction \citep[38\%,][]{lanz10} of the present-day total mass of the cluster, we suggest a second, short ($\Delta t_\mathrm{SF_\mathrm{II}} = 100$~Myr) star-formation episode, after a long (7.5~Gyr) interlude where stars are basically not formed, as inferred from CMD analyses. Hereinafter, we use the I/II subscript when referring to the first/second star-formation episode. Notice that our models give no clues about the physical reasons for this long quiescent phase, and that a very low-level star formation activity in between the two major events \citep[similar to what is inferred for Liller~1, ][]{dale22} could still be accommodated, with no major impact on the model results. In future work we will make use of hydrodynamical simulations \citep{calu19,lacc21} to study in more detail the formation of the young stellar populations.

We start from a configuration that reproduces the chemical properties of the oldest stars of Terzan~5. Model~{\footnotesize T01s} offers a good starting point (see previous section), but it leaves behind only $3.3 \times 10^5$~M$_\odot$ of stars (see Table~\ref{tab:mod-s}, 20$^{\mathrm{th}}$ row), which is too low compared to the current total mass of Terzan~5, namely, $M_{\mathrm{stars}} = (2 \pm 1)\times 10^6$~M$_\odot$ \citep{lanz10}. Thus, we consider a four times more massive progenitor, $M_\mathrm{in} = 4 \times 10^7$~M$_\odot$, yielding a present-day mass $M_\mathrm{stars \, I} = 1.3 \times 10^6$~M$_\odot$\footnote{This value takes into account stripping of 80\% of the long-lived stars with [Fe/H]~$< -0.4$, which allows a better fit to the observed MDF in the low-metallicity regime.} for the sub-solar metallicity component, in excellent agreement with the empirical mass estimate of \citet{lanz10}.

We thus explore three models, namely {\footnotesize S01s}, {\footnotesize S02s}, and {\footnotesize S03s}, that have the same star formation parameters but differ in the amount of gas losses/stellar stripping and in the possibility or not of preferentially retaining SNIa progenitors, as shown in Table~\ref{tab:mod-d}.

\begin{figure}
\centering
\includegraphics[width=21pc]{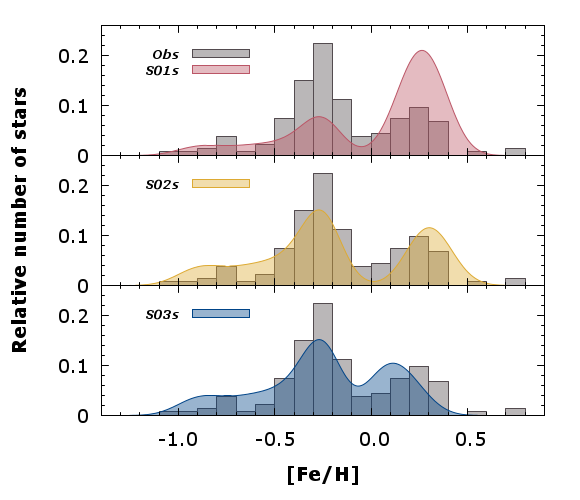}
\caption{Observed \citep[grey histogram,][]{mass14} and predicted (shaded areas) iron distributions of the stellar sub-populations of Terzan~5. The theoretical distributions are convolved with a Gaussian of dispersion $\sigma = 0.1$ dex in order to take the observational errors into account.}
\label{fig:mdf_two}
\end{figure}

\begin{figure}
\centering
\includegraphics[width=21pc]{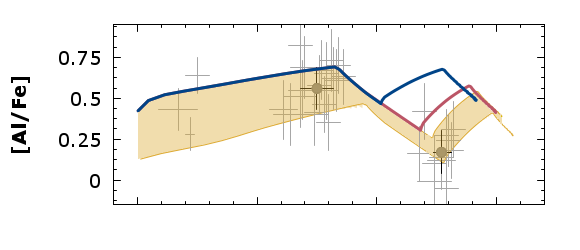}\vspace{-0.72cm}
\includegraphics[width=21pc]{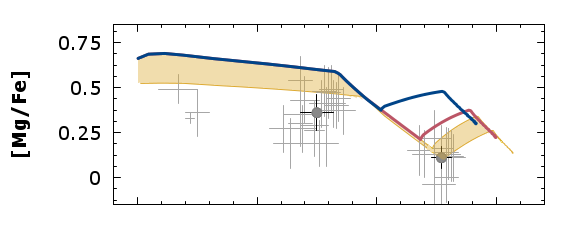}\vspace{-0.72cm}
\includegraphics[width=21pc]{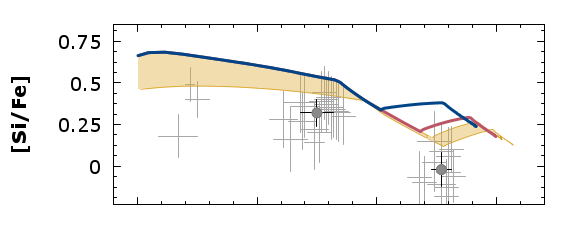}\vspace{-0.72cm}
\includegraphics[width=21pc]{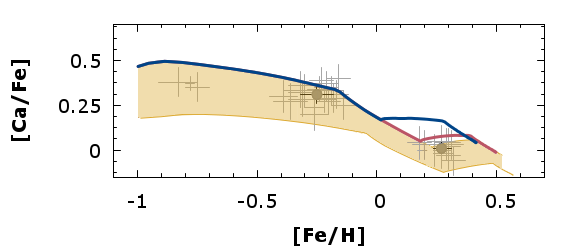}
\caption{\emph{From top to bottom:} [Al/Fe], [Mg/Fe], [Si/Fe], [Ca/Fe] versus [Fe/H] in Terzan~5 predicted by the models shown in Fig.~\ref{fig:mdf_two} (same colour-coding). The shaded areas represent the uncertainties related to the assumed HN fraction (only for model~{\footnotesize S02s}). Data are from \citet{orig11,orig13}. All abundance ratios are normalized assuming the chemical composition of the Sun recommended by \citet{magg22}.}
\label{fig:ele_two}
\end{figure}

Model~{\footnotesize S01s} (see Table~\ref{tab:mod-d} and the shaded red area in Fig.~\ref{fig:mdf_two}, upper panel) makes the young stars forming with the same efficiency of conversion of gas into stars as the old ones (50\%), without any gas loss from the system. This results in the young population dominating the total mass budget, at variance with the empirical mass estimates.

At odds with the previous one, model~{\footnotesize S02s} allows gas losses (see Table~\ref{tab:mod-d}). The removal of a major fraction of the gas leftover from the first star formation episode yields the right proportions of young to old stars, as shown in Table~\ref{tab:mod-d}, last two columns (see also the yellow shaded area in Fig.~\ref{fig:mdf_two}, middle panel). It is worth emphasising that in model~{\footnotesize S02s}, and in model~{\footnotesize S01s} as well, the white dwarfs leading to SNIa explosions that form during the first star formation burst are not lost from the system, but settle in the central regions of the proto-cluster, so that the SNIa rate is boosted in the inner, star-forming regions. Therefore, the youngest stars that are generated in the center reflect the Fe-enriched composition of a medium polluted by all these SNeIa. This well reproduces the observed trends of [Al/Fe], [Mg/Fe], [Si/Fe], and [Ca/Fe] vs [Fe/H] of both the sub-solar and the metal-rich populations (see Fig.~\ref{fig:ele_two}).

Model~{\footnotesize S03s} assumes that most (80\%) of the SNIa precursors are lost together with the low-mass, low-metallicity single stars (Table~\ref{tab:mod-d}). Although the model predictions are still compatible with the observed MDF (blue shaded area in Fig.~\ref{fig:mdf_two}), they fail to reproduce the observed low [Al/Fe] and [$\alpha$/Fe] ratios of the metal-rich component (upper blue line in all panels of Fig.~\ref{fig:ele_two}).

Hence, this analysis shows that gas loss and the retention of a major fraction of Fe produced by SNeIa are fundamental ingredients to reproduce both the MDF and the abundance ratios of the metal-rich population in Terzan~5. As a result, only model {\footnotesize S02s} succeeds in predicting all the observational constraints simultaneously (mass of the two main populations, MDF, and abundance ratios).

\section{Discussion and conclusions}
\label{sec:disc}

Terzan~5 is a complex stellar system located in the outskirts of the inner bulge and characterized by peculiar features that clearly distinguish it from a typical globular cluster. It harbours at least three sub-populations with distinct ages and chemical properties \citep[][and references therein]{ferr16} and the largest population of millisecond pulsars hitherto identified in the MW \citep{rans05,cade18,mart22}. All these characteristics are readily accounted for if Terzan~5 is the compact remnant of a more massive system that self-enriched before losing most of its initial mass. 

Alternative scenarios have been proposed in the literature. The hypothesis of a dwarf galaxy progenitor can be rejected straight away, 
for several reasons. First, the orbit of Terzan~5 clearly points to an object formed \emph{in situ} \citep{mass15,mass19,baum19,call22}. Second, the location in the age-metallicity space of its main stellar population perfectly matches the trend followed by \emph{in-situ} systems (at [Fe/H]~$\sim -0.3$, \emph{ex-situ} objects are more than 2~Gyr younger; see, e.g., \citealt{mass19,krui19}). Finally, it is important to keep in mind that the extreme metallicity regime observed for the two main sub-populations of Terzan~5 is characteristic of the MW bulge environment, while it is clearly incompatible with the iron abundance of any known dwarf galaxy in the local universe.

\citet{mats19} investigated the formation and evolution of young massive clusters during major galaxy mergers. These authors find that several clusters distribute within a few kpc from the centre of the merger remnant (see their figure~4) and that the cluster mass function has an excess around 10$^7$~M$_\odot$ (see their figure~6)\footnote{In this respect, it would be interesting to investigate in simulations if the accretion of the Heracles satellite \citep{hort21} could have triggered the formation of massive clusters in the MW bulge as well.}. Some of the clusters in \citeauthor{mats19}'s (\citeyear{mats19}) simulations have multi-aged populations, which originate from gas capture when the clusters pass through dense gas regions. However, the models of \citep{mats19} predict just a mild metallicity enhancement, certainly not able to reproduce the super-solar population of Terzan~5, likely due to the short time lapse between the main bursts and/or to the neglection of SNeIa. In fact, their second-generation stars are enriched by SNeII only, while the chemical patterns observed in Terzan~5 \citep{orig11} clearly demonstrate that the metal-rich component has been enriched by both Type II and Type Ia SNe.

The possibility that Terzan~5 has been originated by the collision between a GC and a giant molecular cloud of appropriate metallicity has been suggested by several authors \citep{mcke18,bast22}. As discussed by the authors themselves, such events are extremely rare, because they require very fine-tuned combinations of events. However, the increasing number of discovered objects similar to Terzan~5 \citep{ferr21} and the evidence in favor of prolonged star formation histories or more than two star formation bursts (\citealt{dale22}; Crociati et al., in preparation) make these scenarios less likely.

Instead, it is possible that Terzan~5 is the relic of a gaseous clump that originated from the fragmentation of an early disc. Bulges of disc galaxies are known to take their shapes through varied (and not mutually exclusive) formation mechanisms, and the coalescence of sub-systems originating from the fragments of an early, unstable disc is one possibility \citep[e.g.,][]{elme08}. Strong radiative stellar feedback can disperse even the most massive clumps after they turn 5--20\% of their mass into stars in a time-scale of 10--100 Myr, but the stars might remain bound \citep[][see also \citealt{mand17}]{hopk12}.

In this work we explored a set of chemical evolution models aimed at reproducing the mass, the metallicity distribution function (MDF), and the chemical abundance patterns measured so far in Terzan~5, under the assumption that it was the product of the self-enrichment process of one of those primordial gas clumps that could have contributed to the formation of the Galactic bulge at the epoch of the Milky Way bulge assembly. The model that best reproduces the observations is model~{\footnotesize S02s}, which rests on a number of assumptions that will be explored in detail in future work focusing on the dynamical aspects. We summarize our current findings as follows.
\begin{itemize}
\item Among the explored models, model~{\footnotesize S02s} in Table~\ref{tab:mod-d} is the one that allows to reasonably account for all the observables, namely, mass, MDF and abundance ratios of both the old, metal-poor and the young, metal-rich sub-populations.
\item The relative weights of the metal-poor and metal-rich sub-populations of Terzan~5 (comprising about 62\% and 38\% of the current mass of the cluster, respectively) are explained by assuming that Terzan~5 formed via two major star formation episodes separated by a long period (7.5~Gyr) of quiescence or low-level star formation activity, in agreement with CMD analyses. It remains to be seen if delayed energy release from SNeIa allows to keep some gas warm, while still trapped in the potential well of the system, for later star formation.
\item The oldest stars (ages~$\simeq 12$~Gyr) are formed from a gas clump with a mass of a few $10^7$~M$_\odot$. According to our simulations, if the gas clump has a slightly pre-enriched ($Z \simeq 0.005$) chemical composition, a secondary, more metal-poor peak in the MDF is more easily produced.
\item After its formation, the proto-Terzan~5 should lose a large fraction ($>$50\%) of its metal-poor, low-mass stars (owing to mass segregation and tidal stripping) as well as most of the gas left over from the first star formation episode (owing to the combined effects of SN feedback, ram pressure and tidal stripping). In order to explain the chemical properties of the second generation stars, it is crucial to incorporate in the calculations a detailed treatment of the ejecta of the first stellar generations.
\item In particular, the white dwarfs formed during the first star formation burst should settle at the center of the cluster and escape stripping, so that the rate of SNIa explosions is enhanced and Fe production maximized there. This is a necessary condition to reproduce the high-metallicity peak of the MDF and the observed abundance ratios of the metal-rich sub-populations. If this were not the case, the recent star formation burst would tend to restore the abundance ratios to values more typical of massive star ejecta, in disagreement with the observations (see the blue lines in Fig.~\ref{fig:ele_two}).
\item To produce the right proportion of young stars, we need to mix the ejecta of older stellar generations into a fraction of ambient gas. Our model can not discriminate if such gas is coming from cooling of warm gas surrounding the proto-cluster (see the second item above), or if it is collected through interactions with a giant molecular cloud. Hydrodynamical simulations \citep{pfla09,mcke18,mats19} have shown that, in principle, the latter is a viable hypothesis (but see the above discussion). In future work, we will couple the passive evolution of the oldest stellar populations of Terzan~5 with gas accretion via bound cluster-cloud collisions to study self-consistently the rejuvenation of the cluster by means of high-resolution 3D hydrodynamical simulations \citep{calu19,lacc21}.
\end{itemize}

\begin{figure}
\centering
\includegraphics[width=21pc]{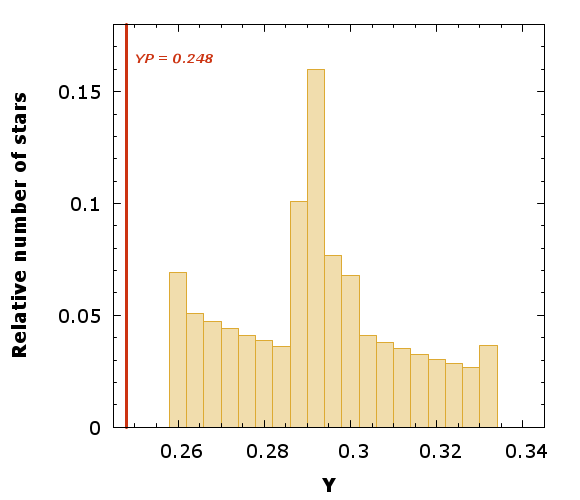}
\caption{Helium distribution of the stellar populations of Terzan~5 after model~{\footnotesize S02s}. The vertical red line indicates the primordial He abundance adopted by our chemical evolution models.}
\label{fig:hedf}
\end{figure}

Finally, we use our most representative two-burst model to make a prediction on the expected He abundance distribution. Model {\footnotesize S02s} (see Sect.~\ref{sec:comp}, Table~\ref{tab:mod-d}, and Figs.~\ref{fig:mdf_two}--\ref{fig:ele_two}) returns a wide He abundance distribution, ranging from $Y \simeq 0.26$ for the oldest stars to $Y \simeq 0.335$ for the youngest ones, with a main peak at $Y = 0.29$ and two local maxima at $Y = 0.26$ and $Y \simeq 0.335$ (see Fig.~\ref{fig:hedf}). This model, like all the others discussed in this paper, starts from a primordial He abundance $Y_\textrm{P} = 0.248$, consistent with the standard model primordial value and with the abundance measured in near-pristine intergalactic gas \citep[see][and references therein]{cook18}. Note that normally stellar isochrones, as those used by \citet{ferr16} or by \citet{bens13} in studying the populations of, respectively, giants and microlensed dwarfs/subgiant bulge stars, adopt a primordial value of $Y_\textrm{P} \simeq 0.24$. Isochrone fitting of the double turn-off in Terzan~5 \citep{ferr16} returns an old age of 12 Gyr, adopting a helium mass fraction $Y = 0.26$, for the dominant, metal-poor component, and an intermediate age of 4.5 Gyr, adopting $Y = 0.29$, for the centrally-concentrated, super-solar metallicity component.

If Terzan~5 is helium-enhanced, as we predict, it will require the development of \emph{ad hoc} isochrones and stellar model atmospheres. Moreover, a self-consistent evaluation of the proto-Terzan~5 chemical enrichment would require the use of stellar yields computed specifically for He-enhanced stars. Unfortunately, the advanced evolution, chemical yields, and final fates of He-rich stars have been explored only in limited ranges of initial stellar mass and metallicity so far \citep{shin15,alth17}.

Another potential problem affecting the yields and, thus, the chemical evolution model predictions, is the neglect of the effects of binary interactions among massive stars, which may be particularly relevant in the high-density environment of the proto-Terzan~5. Investigations of the effects of stellar multiplicity on massive star yields are still very limited, but promising studies are ongoing \citep{farm21,farm23}.


\begin{acknowledgements}
We dedicate this paper to the memory of our colleague and friend, Antonio Luigi Sollima, who sadly passed away earlier this year. We are grateful to the anonymous referee for the constructive feedback on the manuscript. This research was supported by the Munich Institute for Astro, Particle and BioPhysics (MIAPbP) which is funded by the Deutsche Forschungsgemeinschaft (DFG, German Research Foundation) under Germany's Excellence Strategy EXC-2094 390783311. This research is part of the project {\it Cosmic-Lab} at the Physics and Astronomy Department of the University of Bologna (http://www.cosmic-lab.eu/Cosmic-Lab/Home.html). This research has been funded by project {\it Light-on-Dark}, granted by the Italian MIUR through contract PRIN-2017K7REXT (PI: F.~R.~Ferraro). The figures in this article use color-blind friendly palettes retrieved on Paul Tol's website (\url{https://personal.sron.nl/~pault/#fig:scheme_rainbow_discrete}).
\end{acknowledgements}


\bibliography{R23_arxiv} 
\bibliographystyle{aasjournal}



\end{document}